\documentclass[cits]{PoS}

\def\beq{\begin{equation}}
\def\eeq{\end{equation}}
\def\beqn{\begin{eqnarray}}
\def\eeqn{\end{eqnarray}}
\def\nn{\nonumber\\ }

\usepackage{wasysym}   % to have \diameter

\usepackage{wrapfig}
\usepackage{bm}

\title{Studies of eta-mesic nuclei at the LPI electron synchrotron}
\ShortTitle{Studies of eta-mesic nuclei at the LPI electron synchrotron}

\author{%
V.A.~Baskov, A.V.~Koltsov, A.I.~L'vov, A.I.~Lebedev, L.N.~Pavlyuchenko,
\speaker{V.V.~Polyanskiy},
E.V.~Rzhanov, S.S.~Sidorin, G.A.~Sokol\\
P.N. Lebedev Physical Institute, Leninsky prospect 53, Moscow 119991, Russia\\
Email: \email{poly@pluton.lpi.troitsk.ru}}

\author{S.V.~Afanasiev, A.I.~Malakhov\\
Joint Institute for Nuclear Research, Joliot-Curie 6, Dubna
141980, Moscow region, Russia}

\author{A.S.~Ignatov, V.G.~Nedorezov\\
Institute for Nuclear Research, 60-letiya Oktyabrya prospekt 7a,
Moscow 117312, Russia}

\abstract{%
A brief review of searches for $\eta$-mesic nuclei is presented with
emphasis on photoreactions. Results of a new experiment done at the
LPI electron synchrotron are reported. They are as follows.
\\[2ex]
New data on photoproduction of $\eta$-mesic nuclei off $^{12}$C have
been collected at the bremsstrahlung photon beam of $E_{\gamma\;\rm
max} = 850$ MeV. An experimental setup with two plastic time-of flight
spectrometers detected correlated $\pi^+n$ and $pn$ pairs from
annihilation of $\eta$-mesons stopped in the nuclear matter and
measured their velocity distributions. Data analysis was performed
using an Intra Nuclear Cascade Model in the GEANT-3 framework in order
to take into account properties of the setup and physical background.
A separation between charged pions and protons was achieved using
information on velocities and ranges of the particles in plastic
detectors. The obtained data show, apart from previously observed
$\pi^+n$ pairs from one-nucleon annihilation of etas (via $\eta
N\to\pi N$), a presence of emitted correlated $pn$ pairs with
velocities corresponding to the kinematics of the near-threshold
reaction of two-nucleon absorption of the $\eta$-meson in the nucleus
($\eta NN\to NN$). Assuming that such $\pi^+n$ and $pn$ pairs are
mostly produced through formation and decay of quasi-bound states of
the $\eta$-meson and a nucleus (i.e.\ $\eta$-mesic nuclei $_\eta A$),
the cross section of $\eta$-mesic nuclei formation was estimated as
$\sigma(\gamma + {~}^{12}{\rm C}\to{}_\eta A + X) \lesssim 10~\rm \mu b$.}

\FullConference{XXI International Baldin Seminar on High Energy
Physics Problems,\\ September 10-15, 2012\\ JINR, Dubna, Russia}

\begin{document}

\section*{Introduction: $\eta$-mesic nuclei}

$\eta$-mesic nuclei, i.e.\ nuclear systems $_\eta A$ having the
$\eta$-meson bound in a nuclear orbit by strong interaction with $A$
nucleons, have been predicted long ago \cite{haider86,liu86} --- soon
after recognizing the attractive character of the $\eta N$ interaction
at low energies \cite{bhalerao85}. Observations and investigations of
these exotic systems would be very valuable for understanding
meson-baryon interactions in free space and in nuclei and for studies
of properties of hadrons in the dense nuclear matter.

%\begin{figure}[htb]
\begin{wrapfigure}{r}{0.5\textwidth}
\centering\includegraphics[width=0.4\textwidth]{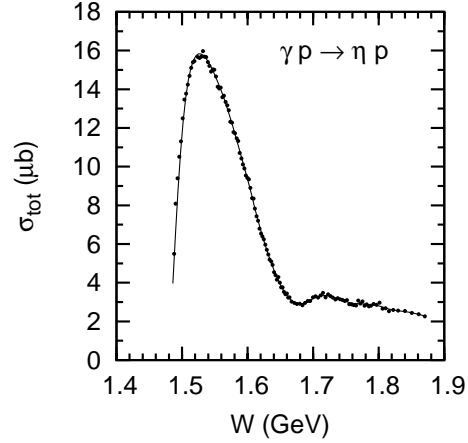}
\caption{Total cross section of $\gamma p\to\eta p$ \cite{mcnicoll10}
as an illustration of the $S_{11}(1535)$ resonance strength in
the $\eta N$ channel.}
\label{eta-x-sect}
\end{wrapfigure}
%\end{figure}
%
The $\eta$-meson, together with pions and kaons, belongs to the SU(3)
octet of pseudoscalar mesons and has, therefore, a similar $q\bar q$
space structure. In contrast to the pion, however, the pseudoscalar
coupling of $\eta$ to the nucleon is empirically rather small
\cite{tiator94}. Nevertheless the amplitude of $\eta N$ $s$-wave
scattering is not as small as that for $\pi N$ scattering because of
the contribution of the $s$-wave resonance $S_{11}(1535)$ which is
actually a chiral partner of the nucleon --- the lowest lying baryon
with the opposite parity to the nucleon. This resonance has the mass
slightly above the $\eta N$ threshold, $m_\eta+m_N = 1486$ MeV, and
owing to its very strong coupling to the $\eta N$ channel [with the
branching ratio ${\rm Br}\;(S_{11}(1535)\to\eta N) \simeq 55\%$] strongly
enhances all interactions in this channel. A nice illustration of this
feature is provided by Mainz data \cite{mcnicoll10} on the total cross
section of $\eta$ photoproduction off protons. A huge near-threshold
enhancement shown in Fig.~\ref{eta-x-sect} is just a manifestation of
the $S_{11}(1535)$ resonance excited in the reaction $\gamma p\to
S_{11}(1535)\to\eta p$.

The $S_{11}(1535)$ resonance strongly contributes to the low-energy
$\eta N$ scattering and, in particular, makes the threshold value of
the $\eta N$ scattering amplitude (i.e.\ the $\eta N$ scattering
length $a_{\eta N}$) positive. In the framework of a dynamical
resonance model for the coupled channels $\pi N$, $\eta N$ and $\pi\pi
N$, Bhalerao and Liu \cite{bhalerao85} found
\beq
   a_{\eta N} = 0.28 + i\, 0.19~\rm fm.
\label{a-etaN-BL}
\eeq
The positive value of ${\rm Re\,}a_{\eta N}$ means an effective
attraction between $\eta$ and $N$, so that one can expect that several
nucleons could jointly bind $\eta$ to a nuclear orbit. The first-order
static-limit on-shell optical potential of $\eta$ in the nuclear
matter at zero energy $E_\eta^{\rm kin}=0$ is equal to
\beq
    U(r) = -2\pi\, a_{\eta N}\,\rho(r) \Big(\frac{1}{m_\eta} + \frac{1}{m_N}\Big),
\eeq
what gives [together with Eq.~(\ref{a-etaN-BL})] $U = -34 -i\;23$ MeV
at normal nuclear matter density $\rho=\rho_0 = 0.17~\rm fm^{-3}$.
The imaginary part of the potential describes a local absorption rate
$\Gamma = -2\,{\rm Im}\,U$ of $\eta$ in the nuclear substance.

With the above strength of the $\eta A$ potential, $\eta$-mesic nuclei
$_\eta A$ are expected to exist for all $A\ge 10$
\cite{haider02,haider09}. Actually, due to a sharp (cusp) energy
dependence of the $\eta N$ scattering amplitude near threshold, Fermi
motion of nucleons and $\eta$ reduces the optical potential
[especially its imaginary part], and this makes $\eta$-mesic nuclei to
exist only for $A\ge 12$. For binding energies and widths of the
lightest $\eta$-mesic nuclei Haider and Liu predicted
\cite{haider02,haider09}
\beqn
   E_\eta = -1.19~{\rm MeV}, && \Gamma_\eta = ~~7.34~{\rm MeV \quad for\quad {}^{12}_{~\eta}C},
\nn
   E_\eta = -3.45~{\rm MeV}, && \Gamma_\eta =  10.76~{\rm MeV \quad for\quad {}^{16}_{~\eta}O},
\nn
   E_\eta = -6.39~{\rm MeV}, && \Gamma_\eta =  13.20~{\rm MeV \quad for\quad {}^{26}_{~\eta}Mg}.
\label{eq:EB-Liu}
\eeqn
Note, however, that a stronger $\eta N$ scattering amplitude was
inferred in some other analyses. For example, using a $K$-matrix model
for coupled channels $\pi N$, $\eta N$, $\gamma N$ and $\pi\pi N$,
Green and Wycech \cite{green97,green05} found from fit to available
data
\beq
   a_{\eta N} = (0.91\pm 0.06) + i\, (0.27\pm 0.02)~\rm fm.
\eeq
With such a big strength of $\eta N$ interaction lighter $\eta$-mesic
nuclei could also exist.

As an example of different predictions for binding energies and widths
of $\eta$-mesic nuclei we mention very elaborated calculations
\cite{oset02a,oset02b,oset02c}, in which a model for meson-baryon
interaction with dynamically generated resonances was build using a
unitarized chiral perturbation theory for coupled channels $\pi N$,
$\eta N$, $K\Lambda$, $K\Sigma$ and $\pi\pi N$ and then self-energies
of all the particles in the nuclear matter were evaluated consistently.
This approach leads to the $\eta N$ scattering length $a_{\eta N} =
0.264 + i\, 0.245~\rm fm$ close to that obtained in
Eq.~(\ref{a-etaN-BL}). The resulting $\eta A$ potential is, however,
found stronger owing to nonlinear dressing effects: $U = -54 -i\, 29$
MeV at normal nuclear density. Also stronger are $\eta$-meson bindings
found in \cite{oset02c}:
\beqn
   E_\eta = -9.71~{\rm MeV}, && \Gamma_\eta = 35.0~{\rm MeV \quad for\quad {}^{12}_{~\eta}C},
\nn
   E_\eta = -12.57~{\rm MeV}, && \Gamma_\eta = 33.4~{\rm MeV \quad for\quad {}^{24}_{~\eta}Mg}.
\label{eq:EB-Oset}
\eeqn

Bindings with equally large widths arise also in calculations
\cite{jido02,nagahiro05,jido08} that use a chiral doublet model and
treat $\eta A$ and $S_{11}(1535)A$ attraction as a result of partial
restoration of chiral symmetry in the dense nuclear matter leading to
reduction of the $S_{11}(1535){-}N$ mass gap. It is clear that
experimental data on energies and widths of $\eta$-mesic nuclei are
needed to test these and many other models and calculations.

\section*{Signature for eta-mesic nuclei produced in photoreactions}

\begin{figure}[htb]
\centering\includegraphics[width=0.8\textwidth]{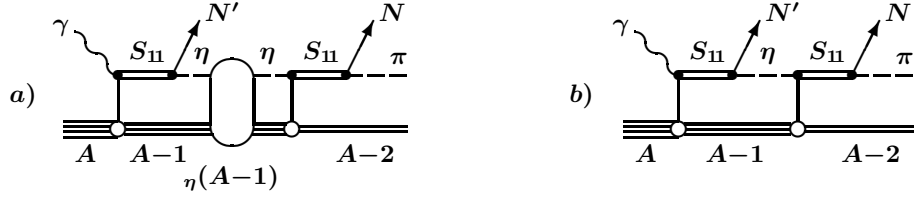}
\caption{a) $\eta$-mesic nuclei formation and decay with the emission
of back-to-back $\pi N$ pairs. b) Background creation of back-to-back
$\pi N$ pairs by unbound $\eta$.}
\label{diagrams-piN}
\end{figure}

A mechanism of $\eta$-mesic nuclei formation and decay in the
photoreaction
\beq
   \gamma + A \to N' + {}_{~\eta}(A-1)\to N' + \pi + N + (A-2)
\label{reac:piN}
\eeq
is shown in Fig.~\ref{diagrams-piN}a. A fast nucleon $N'$ ejected
forward at the first stage of the reaction, i.e.\ in the subprocess
\beq
    \gamma + N' \to N' + \eta_{\rm slow},
\label{reac:eta}
\eeq
escapes the nucleus, whereas a slow $\eta$ is captured by remaining
$A-1$ nucleons to a bound state.  At $E_\gamma \sim 800{-}900$ MeV, a
minimal momentum transfer to $\eta$ in the reaction (\ref{reac:eta})
is not large (less than $70~{\rm MeV}/c$). That is why the total cross
section of $\eta$-mesic nuclei formation off light nuclei (like carbon
or oxygen implied in the following) turns out to be a few $\mu$b
\cite{kohno89,lebedev89,lebedev91,lebedev95,tryasuchev99,tryasuchev01},
i.e.\ $\simeq 2{-}7\%$ of the total cross section $\sigma_{\gamma
A}^\eta$ of inclusive $\eta$ photoproduction, with the exact value
strongly dependent on the assumed strength of the optical potential
$U$.

Energies $E[{}_\eta(A-1)]$ of the produced $\eta$-mesic nuclei can, in
principle, be determined through missing mass measurements in
the reaction $(\gamma,p)$ using tagged photons $\gamma$ and a magnetic
spectrometer for $N'=p$. Indirectly, the same energy
\beq
    E[{}_\eta(A-1)] = E_\eta + E_{A-1} = E_{\pi N} + E_{A-2}
\label{eq:E-eta-(A-1)}
\eeq
can also be found from the observed energy of a correlated
back-to-back $\pi N$ pair produced at the second stage of the reaction
(\ref{reac:piN}) where the captured $\eta$ meson annihilates through
the subprocess
\beq
    \eta N \to N\pi.
\label{etaNpiN}
\eeq
The energy excitation of $(A-2)$ in (\ref{eq:E-eta-(A-1)}) is not a
fixed value. It rather depends on whether an $s$-shell or $p$-shell
nucleon $N$ is knocked out in the process (\ref{etaNpiN}). Therefore a
distribution of the experimental observable $E_{\pi N}$ has
appropriately a bigger width than the width of the $\eta$-mesic
nucleus.

\begin{figure}[h!]
%\vspace{-2ex}
\includegraphics[width=0.33\textwidth]{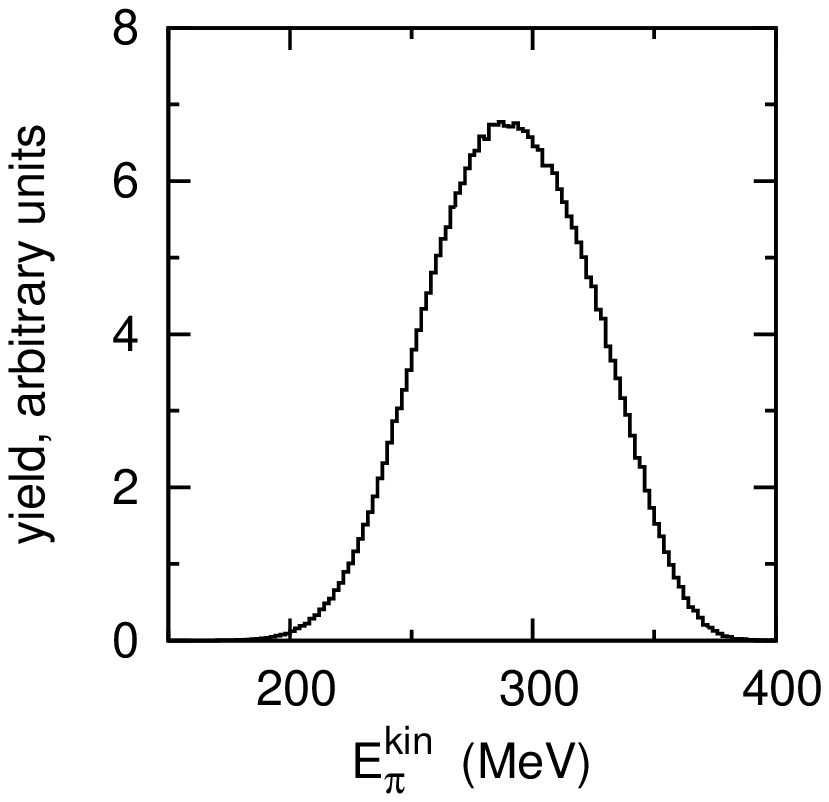}
\includegraphics[width=0.33\textwidth]{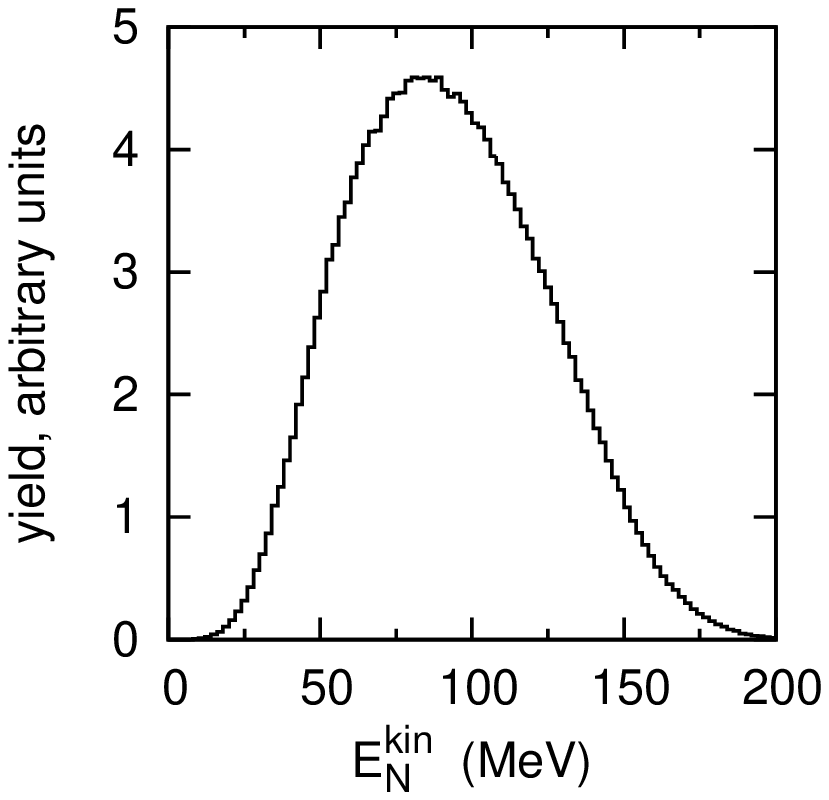}
\includegraphics[width=0.33\textwidth]{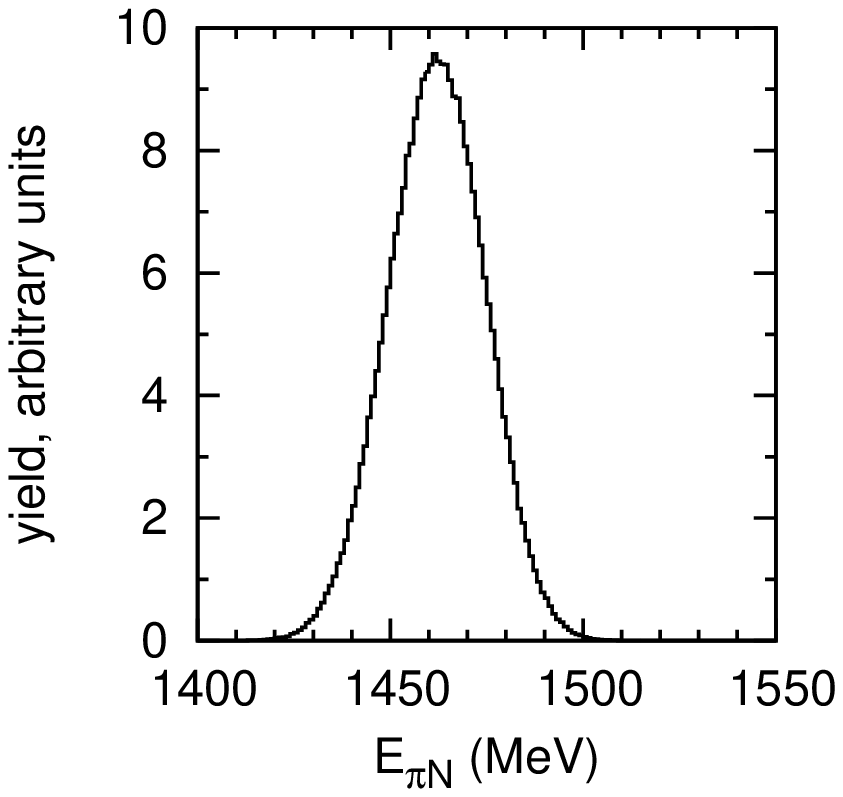}
\par
\includegraphics[width=0.33\textwidth]{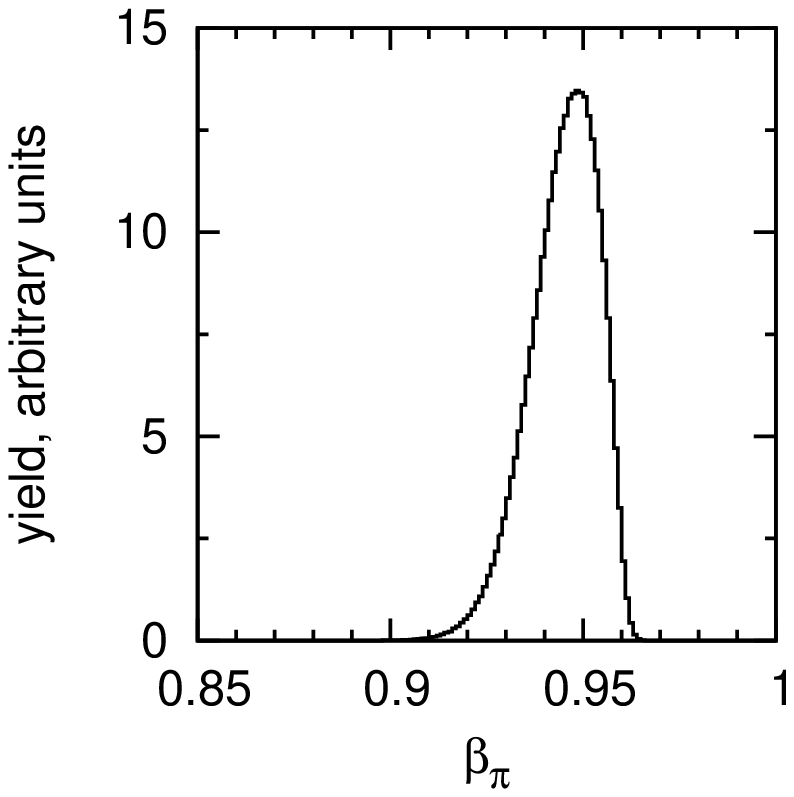}
\includegraphics[width=0.33\textwidth]{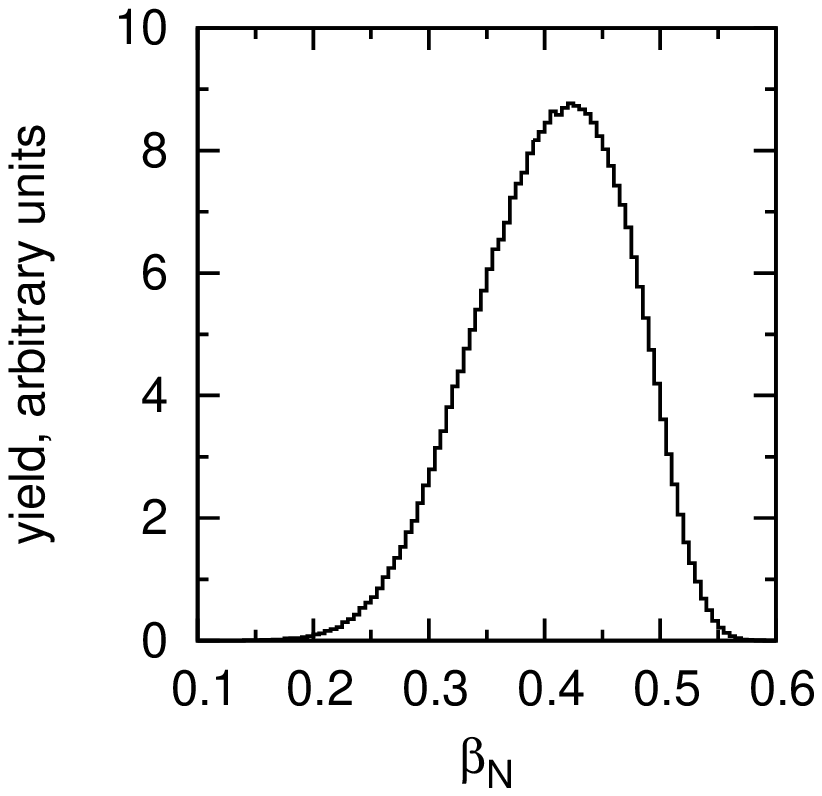}
\includegraphics[width=0.33\textwidth]{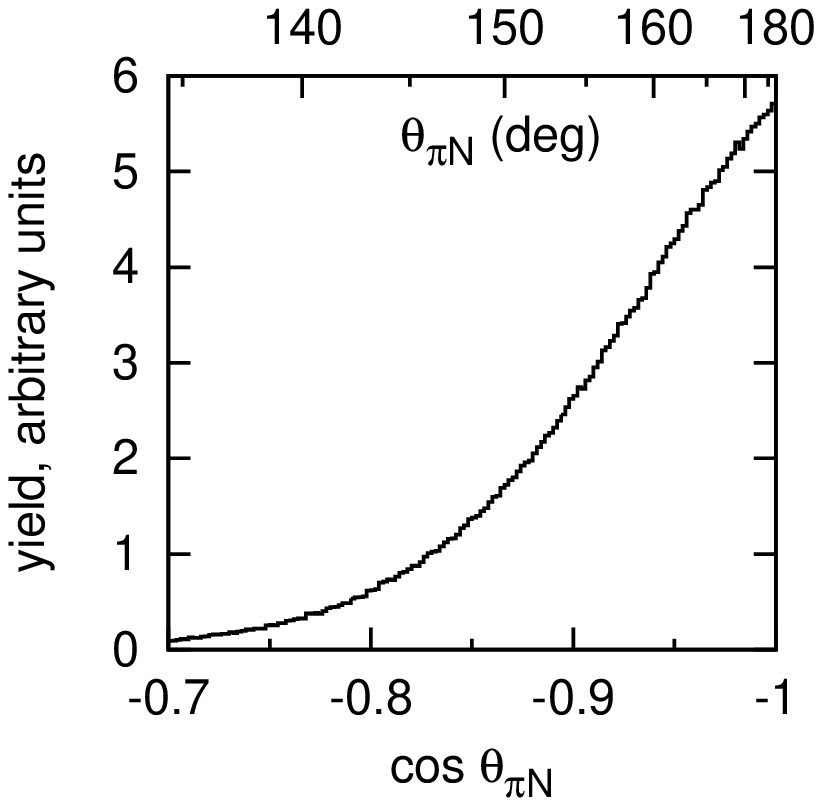}
\vspace{-4ex}
\caption{Simulation of $\pi N$ pairs emitted in $\eta$-mesic nuclei
decays. Shown are distributions over kinetic energies of the
particles, their total energy, velocities, and the $\pi N$ relative angle.}
\label{simulation-piN}
\vspace{3ex}
\end{figure}

Neglecting binding and Fermi motion of nucleons and $\eta$, we have
the following kinematical characteristics of the ejected correlated
$\pi N$ pairs (as for energies, momenta and velocities):
\beqn
    && \sqrt s = E_\pi + E_N = m_\eta+m_N = 1486~\rm MeV,
\nn
    && E_\pi^{\rm kin} = 313~{\rm MeV},\qquad E_N^{\rm kin} = 94~{\rm MeV},
\qquad
       p_\pi = p_N = 431~{\rm MeV}/c,
\nn
    && \beta_\pi = 0.95,\qquad \beta_N=0.42.
\label{kinema-piN}
\eeqn
A simple simulation that takes into account the Fermi motion of
nucleons and $\eta$ as well as binding of these particles reveals that
fluctuations around these ideal parameters are substantial (see
Fig.~\ref{simulation-piN}) [specifically, we used in this simulation
the $\eta$-meson binding energy of 10 MeV with the width 25 MeV; for
nucleons, we assumed a Fermi-gas distribution with binding energies
distributed between 5 and 30 MeV]. In particular, the angle
$\theta_{\pi N}$ between the emitted pion and nucleon may not be so
close to $180^\circ$, and a subtraction of background events with
$\theta_{\pi N} \ne 180^\circ$ used sometimes in practice should be
done cautiously. A shift of the peak down to 1486 MeV in the
distribution of the total energy $E_{\pi N}=E_\pi+E_N$ seen in
Fig.~\ref{simulation-piN} is related with binding of both the
$\eta$-meson (by 10 MeV) and the nucleon (by 15 MeV).

%
%\begin{figure}[htb]
\begin{wrapfigure}{r}{0.4\textwidth}
\vspace{-1ex}
\centering\includegraphics[width=0.35\textwidth]{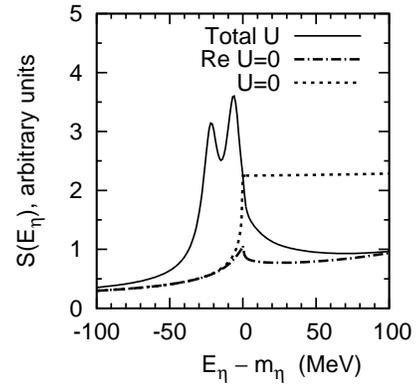}
\caption{Spectral density of $E_\eta$ in a model with a
rectangular-well attractive optical potential $U$ that approximately
simulates the $\eta$-mesic nucleus $^{12}_{~\eta}$C \cite{lvov98}. The
chosen potential was so strong that it bounded $\eta$ both in the $s$
and $p$-wave states. When the attractive potential is turned on, a
pronounced peak (peaks) in the spectral density emerges at
subthreshold resonance energy (energies).}
\label{spectral-function}
\vspace{-3ex}
\end{wrapfigure}
%\end{figure}
%
Notice that $\pi N$ pairs with the characteristics (\ref{kinema-piN})
do not necessary originate from $\eta$-mesic nuclei decays. They can
also be produced by slow etas in a background nonresonance process
shown in Fig.~\ref{diagrams-piN}b. The resonance and nonresonance
processes correspond to a resonance (Breit-Wigner) and nonresonance
part of the full propagator [i.e.\ the Green function $G({\bm r}_1,
{\bm r}_2; E_\eta)$] of the $\eta$-meson moving in the optical
potential $U(r)$. Jointly, these parts generate a complicated spectrum
of $E_\eta$ similar to that obtained in a toy model with a
square-well potential \cite{lvov98,sokol99}. Shown in
Fig.~\ref{spectral-function} is the spectral function in that model,
\beq
   S(E_\eta) = \int \hspace{-1.2ex} \int \rho({\bm r}_1)~ \rho({\bm r}_2)
    ~ |G({\bm r}_1, {\bm r}_2; E_\eta)|^2  \, d{\bm r}_1 \, d{\bm r}_2,
\eeq
that characterizes near-threshold energy distribution of the propagated
etas as well as the near-threshold energy dependence of the yield of
$\pi N$ pairs produced by these $\eta$. Bound states of the
$\eta$-meson give pronounced peaks in the yield of the $\pi N$ pairs
at subthreshold energies $E_\eta$. Generally, observation of a
relatively narrow resonance peak in the spectrum of $E_\eta$ in the
region $E_\eta < m_\eta$ is mandatory for claiming an observation of
$\eta$-mesic nuclei at all. We refer to recent works by Haider and Liu
\cite{haider10a,haider10b} where a deeper and more elaborated
consideration is given in relation with a recent experiment.

Since $\eta$ is isoscalar, the $\pi N$ pairs produced in the
subprocess (\ref{etaNpiN}) have isospin $\frac12$ and hence the
following isotopic contents [for $\eta$-mesic nuclei with $A\gg 1$]:
\beq
  {\rm Br}\,(\pi N) =\quad\left\{\begin{array}{lll}
  1/3 & {\rm for} & \pi^+n,\\
  1/6 & {\rm for} & \pi^0p,\\
  1/6 & {\rm for} & \pi^0n,\\
  1/3 & {\rm for} & \pi^-p.
\end{array}\right.
\label{piN-modes}
\eeq
From these, the channel $\pi^+n$ was chosen for detection in our
experiment.

\section*{Previous searches for $\eta$-mesic nuclei}

Searches for $\eta$-mesic nuclei began very soon after their
predictions \cite{haider86} followed by suggestions
\cite{liu86,kohno89,lebedev89,lebedev91,kohno90} to seek these novel
high-energy nuclear excitations in missing-mass experiments using the
inclusive reactions $(\pi^+,p)$ and $(\gamma,p)$.

The first two experiments have been done along this line in 1988 at
Brookhaven \cite{chrien88} and Los Alamos \cite{lieb88a,lieb88b}. In
both experiments, a $\pi^+$ beam was used and several targets (Li, C, O
and Al) were examined. The inclusive $(\pi^+,p)$ reaction
\beq
   \pi^+ + A \to ~{}_\eta(A-1) + p
\eeq
was studied in \cite{chrien88} with a magnetic spectrometer, whereas
the Los Alamos experiment had also an additional $4\pi$ BGO crystal
ball for detecting charged paticles ejected in the subprocess
(\ref{etaNpiN}) of $\eta$-mesic nuclei decays to $\pi N$ pairs in
coincidence with the forward proton $p$.

The Brookhaven experiment did not find a theoretically expected signal
\cite{liu86} --- a relatively narrow peak of a predicted strength in
the missing mass spectrum. The team working at Los Alamos did report a
preliminary evidence for a wanted peak for the $^{16}$O target but
this report was not confirmed (published) since then.

It was recognized in the following that the above obtained negative or
incomplete results do not necessarily mean that the predicted
$\eta$-mesic nuclei do not exist. It was possible that the binding
energies and especially the widths of the $\eta$ bound states were
theoretically underestimated. This point of view was supported by
many-body calculations \cite{chiang91} taking into account some
effects disregarded in the first theoretical works
\cite{haider86,liu86}, in particular --- dressing, binding and
collisional decays of the $S_{11}(1535)$ resonance in the dense nuclear matter.
The analysis of \cite{chiang91} was later extended and revised
\cite{oset02a,oset02b,oset02c} (in particular, dressing of mesons was
also included) with the main conclusion survived that $\eta$-mesic
nuclei widths are bigger than those found in \cite{haider86,liu86}.

The next experiment has been performed at the Lebedev Physical
Institute in Moscow/Troitsk \cite{sokol99,sokol00} (see also a summary
in \cite{sokol08}). It was triggered \cite{sokol94,lebedev95a} by a
suggestion \cite{sokol91} to seek $\eta$-mesic nuclei through
observing decay products of $\eta$-mesic nuclei, namely two correlated
back-to-back particles, a pion and a nucleon, ejected in the process
of annihilation of captured $\eta$-mesons in the nucleus,
Eq.~(\ref{etaNpiN}). It was hoped that a background for the two very
energetic particles (the pion and the nucleon) ejected in decays of
$\eta$-mesic nuclei transversely to the beam would be lower than that
for ejection of forward protons in the inclusive processes. Besides,
it was hoped that background conditions in photon-induced reactions
would be generally better than those in pion-induced ones.

Studies of the reaction
\beq
   \gamma + {}^{12}{\rm C}\to
     ({}_{~\eta}^{11}{\rm Be} {\rm ~~or~~} {}_{~\eta}^{11}{\rm C}) + N
     \to \pi^+ + n + X + N
\label{reac-C:piN}
\eeq
done in the middle of 1990's at the LPI electron synchrotron indeed
showed a signal of an enhanced production of the correlated
back-to-back $\pi^+n$ pairs ejected transversely to the photon beam
when the photon energy exceeded the $\eta$-meson photoproduction
threshold. Energy resolution of the experimental setup was, however,
not sufficient to resolve a peak similar to that shown in
Fig.~\ref{spectral-function} and to determine whether the observed
correlated pairs were produced by bound or unbound intermediate etas.

After the works \cite{sokol99,sokol00} gaining and using
information on the decay products became mandatory for experiment
planning and data analysis in all further searches for
$\eta$-mesic nuclei.

In 2004 an evidence for the $\eta$-mesic nucleus $_\eta^3$He formed in
the reaction
\beq
  \gamma + {}^3{\rm He} \to {}_\eta^3{\rm He} \to \pi^0 + p + X
\eeq
has been reported from Mainz \cite{pfeiffer04}. A resonance-like
structure was observed in a contribution to the cross section from
back-to-back $\pi^0p$ pairs found after a background subtraction. A
later study \cite{pheron12} revealed, however, that the background has
a rather complicated structure, so that the conclusions of
Ref.~\cite{pfeiffer04} cannot be confirmed. At the moment their
statement is that the existence of the $\eta$-mesic nucleus
$_\eta^3$He is not yet established.

One more attempt to find $\eta$-mesic nuclei by detecting their
$\pi^-p$ decay products has recently been done at the JINR nuclotron
\cite{afanasiev11}. The reaction studied was
\beq
   d + {}^{12}{\rm C}\to
     ({}_{~\eta}^{11}{\rm Be} {\rm ~~or~~} {}_{~\eta}^{11}{\rm C}) + N_1 + N_2
     \to \pi^- + p\, + X + N_1 + N_2.
\label{reac:Dubna}
\eeq
The measured effective mass spectra of the correlated back-to-back
$\pi^-p$ pairs show a presence of resonance-like peaks lying slightly
below the threshold energy $m_\eta+m_N=1486$ MeV. However, a
consistent interpretation of these peaks was not yet obtained.

To date the strongest evidence for the existence of $\eta$-mesic
nuclei came from the precision COSY-GEM experiment \cite{budzanowski09}.
Following ideas of the work \cite{hayano99} borrowed in turn from
previous experience in studying deeply-bound pionic states in nuclei,
the reaction
\beq
  p + {}^{27}\!{\rm Al} \to {}^3{\rm He} + {}^{25}_{~\eta}{\rm Mg}
        \to {}^3{\rm He} + p + \pi^- + X
\label{reac:COSY-GEM}
\eeq
of a recoilless formation of the $\eta$-mesic nuclei $^{25}_{~\eta}\rm
Mg$ was explored and the mass of this $\eta$-mesic nucleus was
determined through precision missing-mass measurements in $(p,
{}^3{\rm He})$. A clear peak was found in the missing mass spectrum
that corresponds to the binding energy $-13.13\pm 1.64$ MeV and the
width $10.22\pm 2.98$ MeV of the formed $\eta$-mesic nucleus. An upper
limit of $\approx 0.5$~nb was found for the cross section of the
$\eta$-mesic nucleus formation.

Recently Haider and Liu argued \cite{haider10a,haider10b} that the
observed peak in (\ref{reac:COSY-GEM}) is shifted down from the
genuine binding energy of $\eta$ because of interference of the
resonance and nonresonance mechanisms of the reaction (similar to
those shown in Fig.~\ref{diagrams-piN}). This very interesting effect
signifies that the genuine $\eta$ binding in ${}_{~\eta}^{25}{\rm Mg}$ is~
$\approx -8$ MeV with the width $\approx 19$ MeV.

\section*{On the two-nucleon decay mode of $\eta$-mesic nuclei}

%
%\begin{figure}[htb]
\begin{wrapfigure}{r}{0.4\textwidth}
\centering\includegraphics[width=0.35\textwidth]{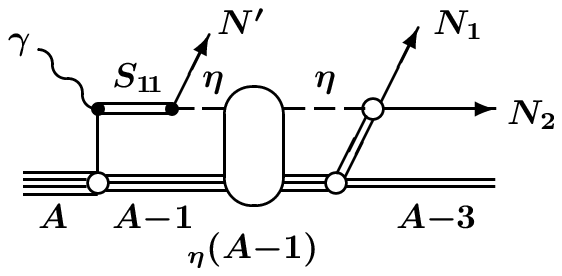}
\caption{$\eta$-mesic nucleus formation and decay with emission of a
back-to-back $NN$ pair.}
\label{diagram-NN}
\vspace{5ex}
%\end{wrapfigure}
%\end{figure}
%
%\begin{figure}[htb]
%\begin{wrapfigure}{r}{0.4\textwidth}
\vspace{2ex}
\centering\includegraphics[width=0.4\textwidth]{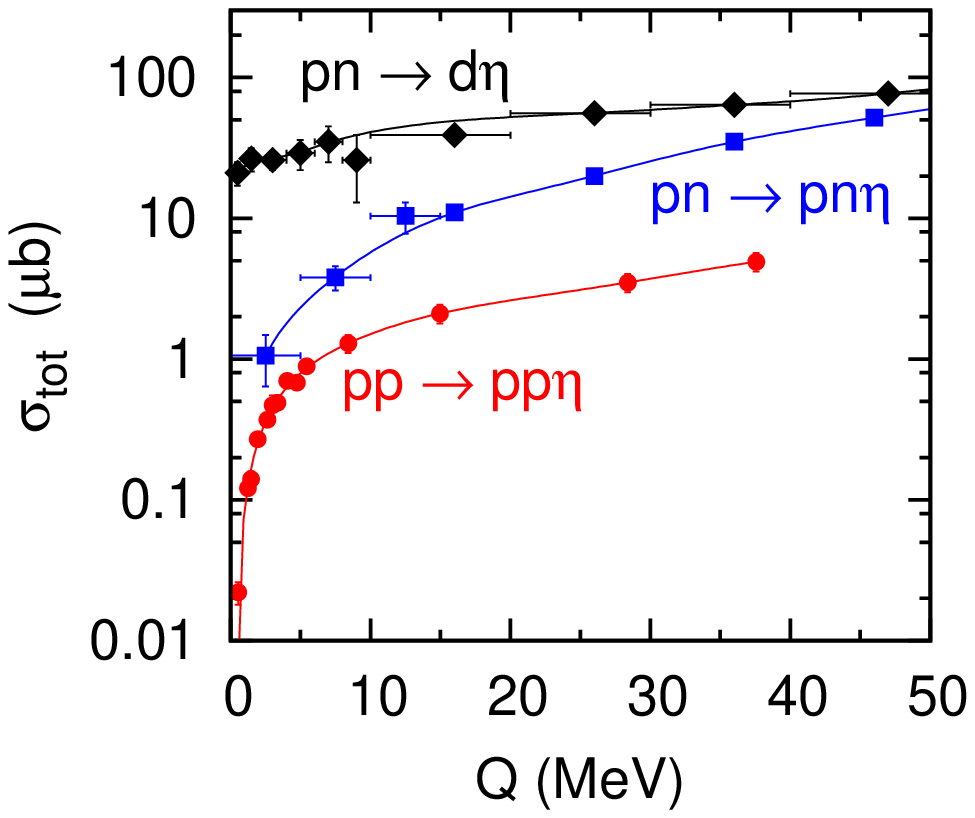}
\caption{Uppsala-Celsius \cite{calen96,calen97,calen98a,calen98b} and
COSY \cite{smyrski00,moskal09} data (with statistical errors only) on
the total cross sections of $pp\to pp\eta$, $pn\to pn\eta$ and $pn\to
d\eta$ near threshold vs the excess energy $Q=\sqrt s - 2m_N -
m_\eta$.}
\label{fig-etaNN}
\vspace{2ex}
\end{wrapfigure}
%\end{figure}
%
The main novelty in our present research is exploring a new
possibility for searching for $\eta$-mesic nuclei, namely through
observation of their two-nucleon decay mode arising owing to the
two-nucleon absorption of the captured $\eta$ in the nucleus,
\beq
  \eta NN \to NN,
\label{2N-absorption}
\eeq
see Fig.~\ref{diagram-NN}.
Ejected in this process correlated back-to-back nucleons of the $NN$
pairs have very high energies ($E_N^{\rm kin}\simeq \frac12 m_\eta = 274$ MeV) and
momenta ($p_N\simeq 770~{\rm MeV}/c$), so that they are to be visible
(especially in coincidence) at the background of other particles
emitted in photoreactions at $E_\gamma\sim 800$ MeV and thus should
provide a bright signature for the $\eta$-mesic nucleus formation.

The $NN$ pair production in decays of $\eta$ in the nuclear matter was
considered among other channels by Chiang, Oset and Liu
\cite{chiang91} in terms of the self-energy of $S_{11}(1535)$ that includes a
contribution of $S_{11}(1535)N\to NN$. A more direct and rather transparent
evaluation of this process has been done by Kulpa and Wycech \cite{kulpa98b} who used
available experimental data on the inverse reactions $pp\to pp\eta$,
$pn\to pn\eta$ and $pn\to d\eta$ and then converted them into the rate
of (\ref{2N-absorption}). In terms of the imaginary part $W_{NN}$ of
the optical potential $U$, this rate was found to be proportional to
$\rho^2$, being $W_{NN}=3.4$ MeV at central nuclear density. This is
only about 15\% of $W_N \sim 23$ MeV related with the absorption of
$\eta$ by one nucleon. Nevertheless such a small fraction of $NN$ can
be quite visible experimentally because of a specific isotopic
contents of the $\pi N$ and $NN$ pairs.

\begin{figure}[htb]
\hspace{-4ex}
\includegraphics[width=0.5\textwidth,height=13ex]{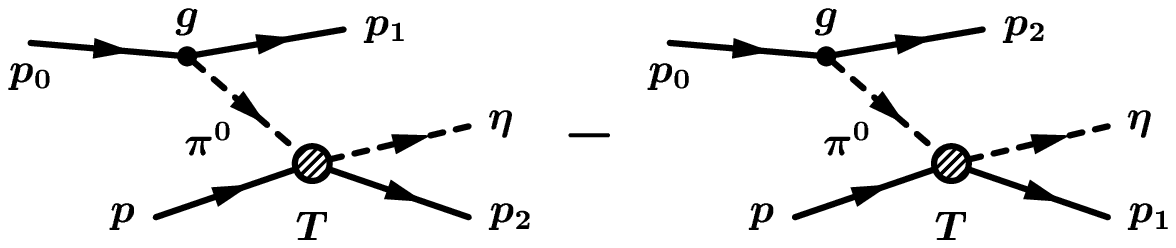}
\hspace{5ex}
\includegraphics[width=0.5\textwidth,height=13ex]{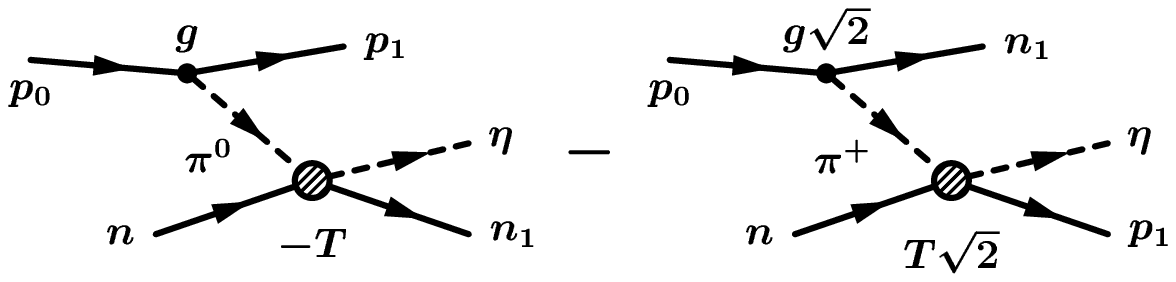}
\caption{Pion-exchange mechanism of $NN\to NN\eta$. Isospin factors,
which accompany the $\pi NN$ coupling $g$ and the $\pi N\to\eta N$
amplitude $T$, and the Fermi-statistics signs (both shown in this
Figure) jointly determine the big difference between the cross
sections of $pp\to pp\eta$ and $pn\to pn\eta$ (plus $pn\to d\eta$).
Antisymmetrization of the initial state and initial/final state
interactions are not shown.}
\label{diagrams-NNeta}
\vspace{4ex}
\end{figure}

The matter is that $\gtrsim 90\%$ of these $NN$ pairs are proton plus
neutron because the cross section of $pp\to pp\eta$ (and $nn\to
nn\eta$) is by order or magnitude less than that of $pn\to pn\eta$
(plus $pn\to d\eta$), see Fig.~\ref{fig-etaNN} where pertinent
Uppsala-Celsius \cite{calen96,calen97,calen98a,calen98b} and COSY
\cite{smyrski00,moskal09} data are shown (and see also, e.g.,
\cite{baru03} for theoretical explanations). This difference can be
traced to isospin factors and Fermi statistics signs in the dominating
pion-exchange mechanism of the reaction $NN\to NN\eta$ shown in
Fig.~\ref{diagrams-NNeta}. If the experimental setup detects one
charged and one neutral particle from the pairs, it detects $\sim90\%$
of $NN$ and only $\sim33\%$ of $\pi N$. Then count rates of the setup
would not be so different for $pn$ and $\pi^+n$ pairs. That seems to
be exactly what we see in our experiment.

\begin{figure}[h!]
%\vspace{4ex}
\centering
\includegraphics[width=0.36\textwidth]{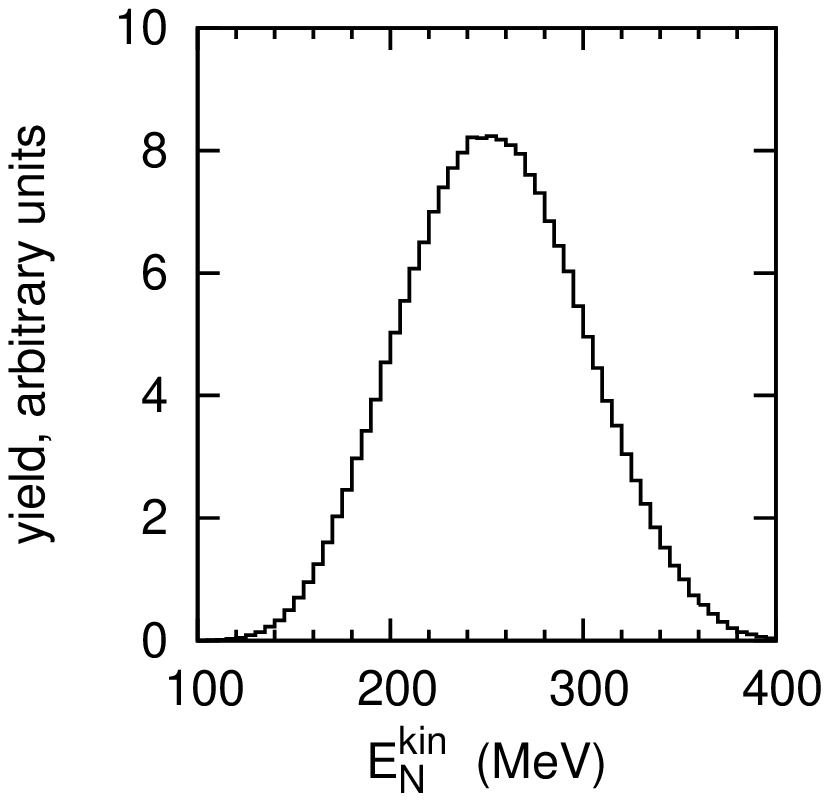}
\includegraphics[width=0.36\textwidth]{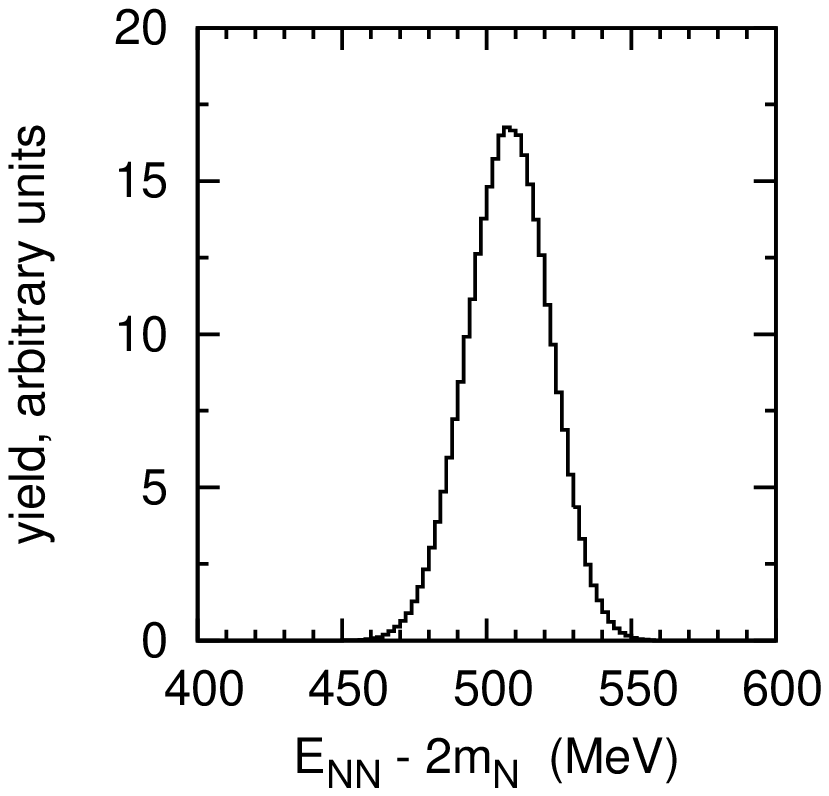}
\\
\includegraphics[width=0.36\textwidth]{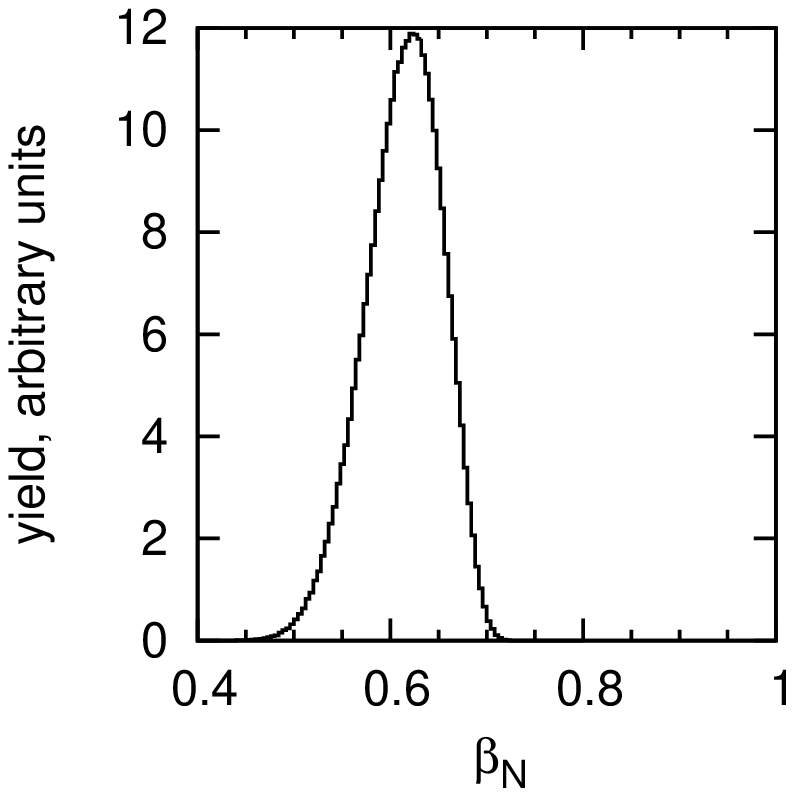}
\includegraphics[width=0.36\textwidth]{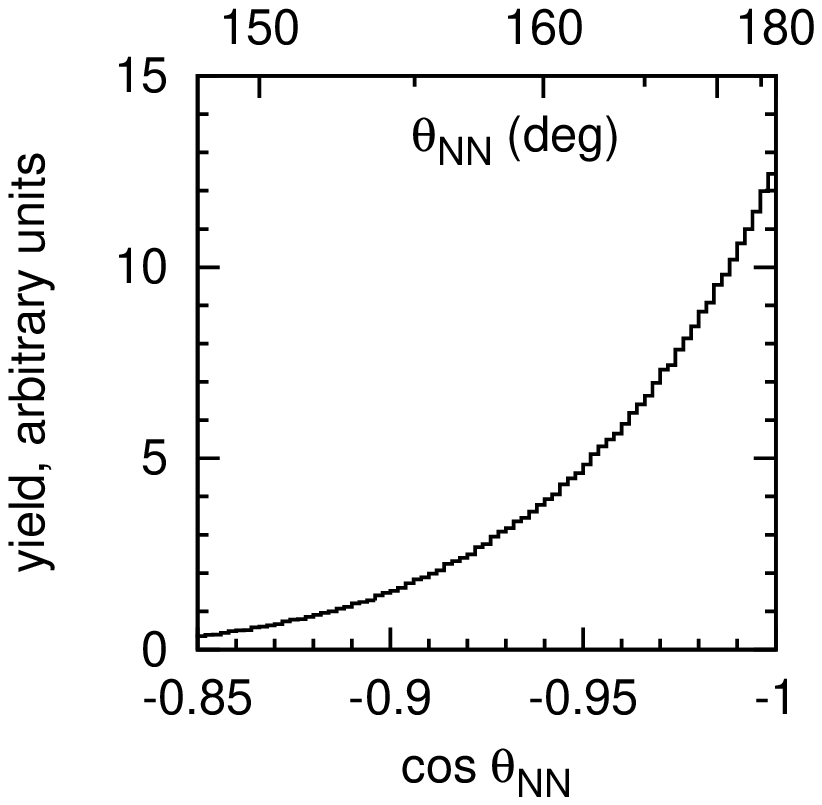}
\vspace{-2ex}
\caption{Simulation of $NN$ pairs emitted in decays of $\eta$-mesic
nuclei. Shown are distributions over the kinetic energy and velocity
of one of the nucleons, the total energy of the pair and the relative angle.}
\label{simulation-NN}
\vspace{2ex}
\end{figure}

Neglecting binding effects and effects of Fermi motion of nucleons and
$\eta$, we have the following kinematical characteristics of the
correlated $NN$ pairs (i.e.\ energies, momenta, velocities) ejected in
$\eta$-mesic nuclei decays:
\beq
     E_{N_1}^{\rm kin} = E_{N_2}^{\rm kin} = \frac12 m_\eta = 274~{\rm MeV},
\qquad
       p_{N_1} = p_{N_2} = 767~{\rm MeV}/c,
\qquad
      \beta_{N_1} = \beta_{N_2} = 0.63.
\label{kinema-NN}
\eeq
Actually, the Fermi motion and binding leads to fluctuations around
these ideal parameters as a simple simulation reveals, see
Fig.~\ref{simulation-NN}. Note that the angular correlation in the
$NN$ pairs is stronger than that in the $\pi N$ pairs --- owing to
higher momenta of particles in $NN$.

The first studies of the photoreaction
\beq
   \gamma + {}^{12}{\rm C}\to
     ({}_{~\eta}^{11}{\rm Be} {\rm ~~or~~} {}_{~\eta}^{11}{\rm C}) + N
     \to p + n + X + N
\label{reac:NN}
\eeq
have recently been done at the LPI synchrotron and we report below on
the obtained results.

\section*{Experimental setup at LPI}

%
%\begin{figure}[htb]
\begin{wrapfigure}{r}{0.3\textwidth}
\vspace{4ex}
\centering\includegraphics[width=0.3\textwidth]{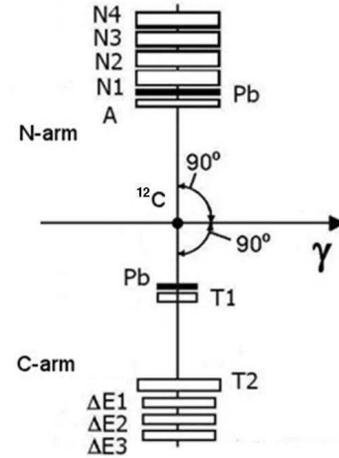}
\caption{Layout of the experimental setup at LPI.}
\label{exp-setup}
%\vspace{-4ex}
\end{wrapfigure}
%\end{figure}
%
Our experiment was performed at the bremsstrahlung photon beam of the
1.2-GeV electron synchrotron of the Lebedev Physical Institute. Photons
were produced with an electron beam of intensity $I_e \simeq 10^{12}
~\rm s^{-1}$ and the duty factor $\simeq 10\%$. The energy of the
beam was usually $E_e = E_{\gamma\;\rm max} = 850~\rm MeV$ (i.e. above
the $\eta$ photoproduction threshold off free nucleons, $E_{\eta\;\rm
thr}=708~\rm MeV$); additional measurements of subthreshold
backgrounds have been done at $E_e = E_{\gamma\;\rm max} = 650~\rm
MeV$.

The experimental setup included two time-of-flight arms (two
scintillation telescopes --- C and N arms) for detecting in
coincidence charged and neutral particles (back-to-back pairs), see
Fig.~\ref{exp-setup}.
%(and also Fig.~\ref{exp-setup-live}).
These arms were both positioned at $90^\circ{-}90^\circ$ with respect
to the beam axis in order to minimize background.

%\begin{figure}[htb]
%\vspace{0ex}
%\centering\includegraphics[width=0.8\textwidth]{fig-setup-live.eps}
%\caption{Experimental setup at LPI.}
%\label{exp-setup-live}
%\end{figure}

The C-arm used for detection of charged particles is a plastic TOF
spectrometer for charged pions and protons. It consists of a start
detector T1 ($20\times 20\times 2~\rm cm^3$), a stop detector T2
($50\times 50\times 5~\rm cm^3$) and three energy losses detectors
$\Delta E_1$, $\Delta E_2$ and $\Delta E_3$ ($40\times 40\times 2~\rm
cm^3$ each). A 4~mm lead (Pb) plate was used in some runs for TOF
calibrations with ultrarelativistic electrons/positrons produced in
the lead plate with high-energy photons emitted from the target owing
to production and decays of neutral pions.

The N-arm is a plastic TOF spectrometer for neutrons. It consists of a
veto counter A ($50\times 50\times 2~\rm cm^3$) and four plastic
blocks --- the neutron detectors N1, N2, N3 and N4 ($50\times 50\times
10~\rm cm^3$ each). Again, a 4~mm lead plate was used in some runs for
TOF calibrations. The efficiency of the N-arm for neutrons of energies
above 50 MeV was $\approx 30\%$.

In both arms each volume of scintillator counters/blocks was
viewed from corners with 4 phototubes. The time-of-flight bases in
the C and N arms were 1.4~m and the time resolution was $\simeq
200$~ps ($1\sigma$). The target was a carbon cylinder of the 10~cm
length along the beam axis. Its diameter was 4~cm, i.e.\ slightly
more than the diameter of the collimated photon beam (3~cm). The
distance between the target and the start detector T1 was 0.7~m.

Mostly, the setup was the same as in our previous work
\cite{sokol00,sokol08} but a few useful changes have been made:
\begin{itemize}
\item  $\Delta E_i$ detectors have been placed after the time-of-flight
interval T1-T2. This enabled us to have a better $\pi^\pm/p$ separation and
time resolution.
\item  A transverse size of the start detector T1 was cut off
according to required geometry. This reduced a background load of the
C-arm.
\item  A thickness of the start detector was also reduced in order to
improve time resolution.
\item  All unnecessary layers of absorbers used previously to suppress
radiative backgrounds have been removed from the time-of-flight
interval, with the effect of reducing the $e^+/e^-$ background created
by photons from $\pi^0$ decays.
\end{itemize}

%\begin{figure}[htb]
\begin{wrapfigure}{r}{0.4\textwidth}
\vspace{-2ex}
\centering\includegraphics[width=0.35\textwidth]{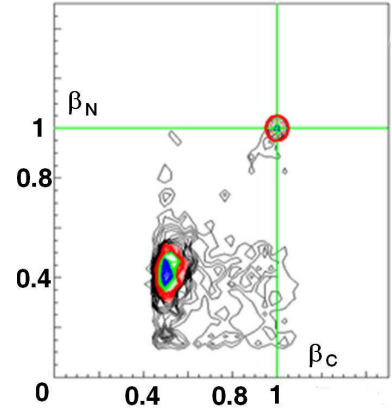}
\caption{A two-dimensional plot of particle velocities, $\beta_C$ and
$\beta_N$, in the C and N arms.}
\label{2-dim-bb}
\end{wrapfigure}
%\end{figure}
General tests of the setup, including preliminary time calibrations of
the arms, have been done in special runs, in which the quasifree
reaction $\gamma p\to\pi^+n$ inside carbon nuclei was observed. In
such runs the two arms of the setup have been moved to the angles
$50^\circ{-}50^\circ$ where the high count rate enabled one to do the
calibrations quickly. Lead convertors used in these runs provided
reliable ultrarelativistic reference points $\beta=1$ for particle's
velocities $\beta_C$ and $\beta_N$ measured in the C- and N-arms. A
two-dimensional $\beta_C{-}\beta_N$ plot on Fig.~\ref{2-dim-bb}
illustrates this procedure.

The calibration done provided a linear scale of velocities in the
range $\beta  = 0.6{-}1$ with errors of about 3\% ($1\sigma$). We have
checked the linearity of the scale by using cosmic rays and setting
different distances between detectors.

\section*{Results and comparison with simulations}

Measurement runs have mostly been done in 2009 at two maximal beam
energies: $E_{\gamma\;\rm max} =  650$ MeV and 850 MeV. The on-line
trigger was a coincidence of particles in the C- and N-arms within a
time gate of 50~ns.

For further off-line analysis events were selected with an additional
condition of sufficiently long ranges of the charged particles,
\beq
   \Delta E_i > E_i^{\rm thr} \quad \mbox{for \underline{~all} ~$i=1,2,3$}
\label{eq:selection} 
\eeq 
with experimentally adjusted thresholds $E_i^{\rm thr}$. In this way
low-energy particles in the C-arm were rejected.

\begin{figure}[htb]
\vspace{5ex}
\centering\includegraphics[width=0.6\textwidth]{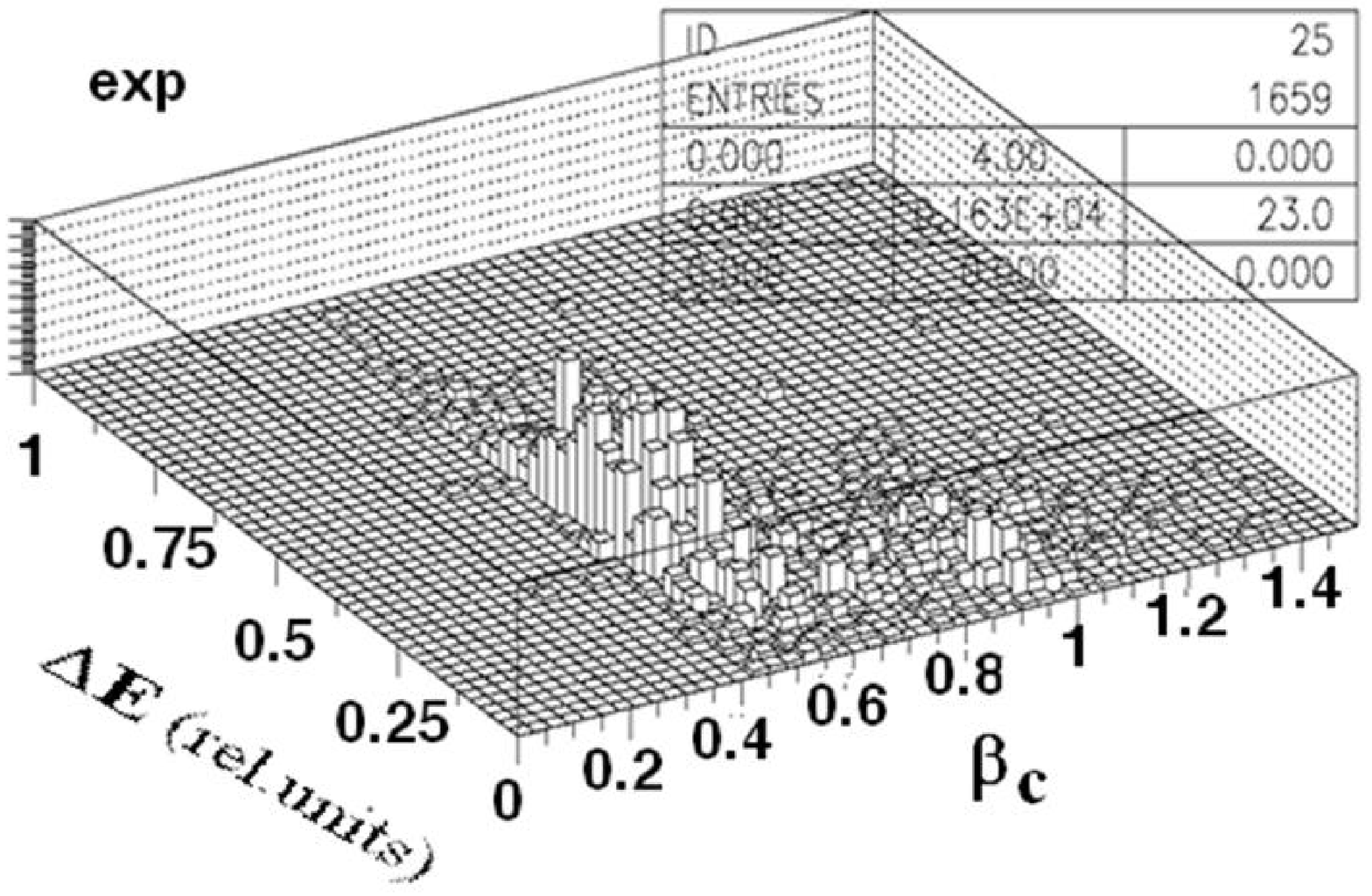}
\caption{Two-dimensional $\Delta E{-}\beta_C$ distribution, experiment.}
\label{2-dim-bE}
\par
\vspace{5ex}
\centering\includegraphics[width=0.6\textwidth]{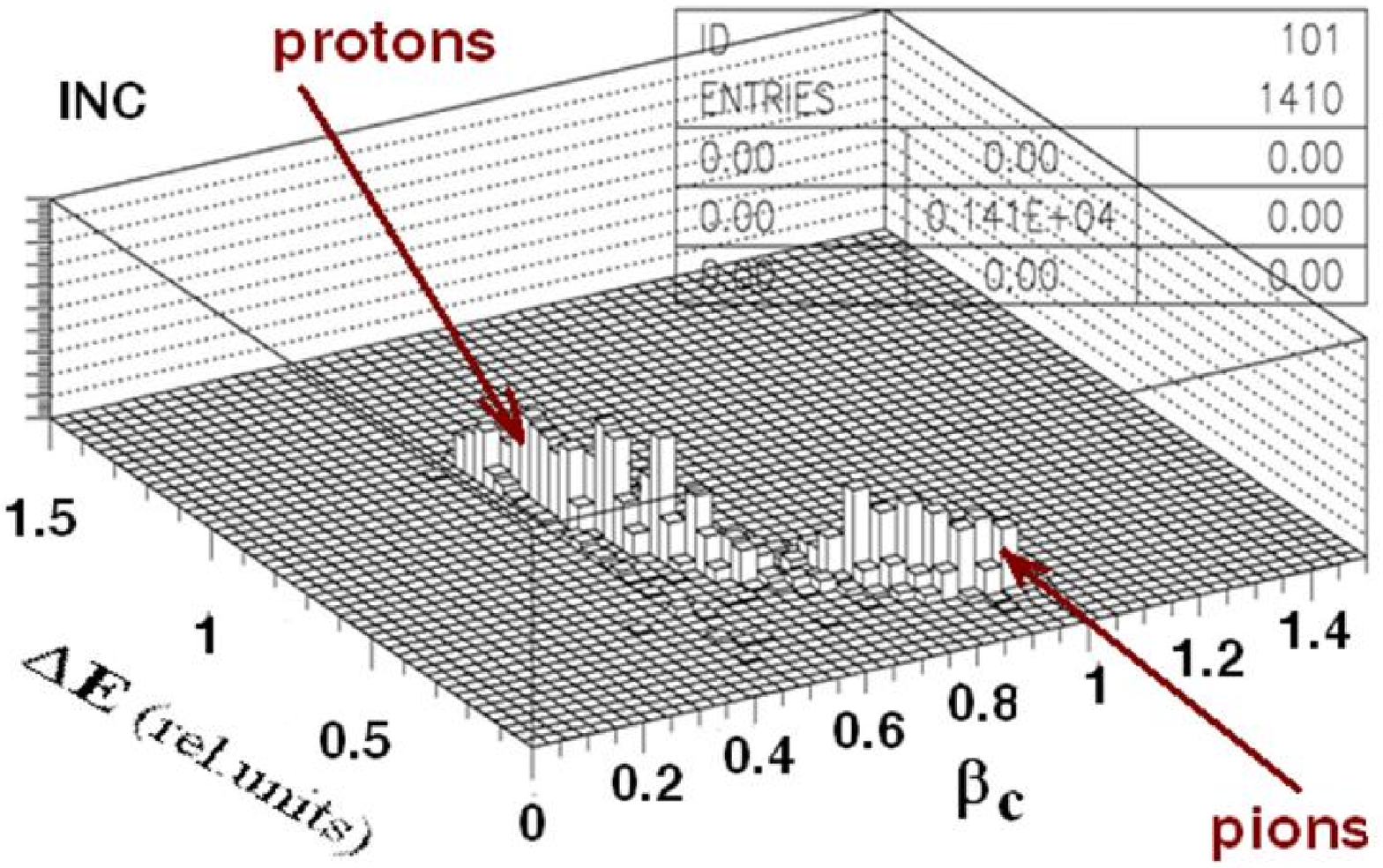}
\caption{Two-dimensional $\Delta E{-}\beta_C$ distribution, the INC model.}
\label{2-dim-bE-INC}
\end{figure}

A two-dimensional histogram in the variables $\Delta E{-}\beta_C$\,,
where $\Delta E$ is the minimal energy loss in the $\Delta E_i$
detectors,
\beq
   \Delta E=\min_i(\Delta E_i),
\eeq
is shown in Fig.~\ref{2-dim-bE} for the beam energy $E_{\gamma\;\rm
max} = 850$ MeV. Results of simulations using the Intra Nuclear
Cascade (INC) model \cite{pschenichnov97} in the GEANT-3 framework are
shown in Fig.~\ref{2-dim-bE-INC} for comparison. The INC model takes
into account production of various mesons and baryon resonances, their
free propagation in the nuclear matter, and then various $2\to 2$
collisional reactions including $\eta N\to\pi N$. This model
successfully describes many photoreactions in wide kinematical ranges
as was demonstrated, beyond \cite{pschenichnov97}, in simulations of
the GRAAL experiment at energies 500--1500 MeV \cite{ignatov08}.
Binding effects for $\eta$ and reactions like $\eta NN\to NN$ were not
included into the model, so one can try to find effects arising due to
formation and decay of $\eta$-mesic nuclei through characteristic
deviations of the model predictions from the experimental data.

The simulation shows that the selection (\ref{eq:selection}) of
particles with sufficiently long ranges distinguishes very well protons
(as particles with $\beta_C \leq 0.7$) and pions (as particles with
$\beta_C \geq 0.7$): the overlap is less than 1\%.

\begin{figure}[htb]
\vspace{3ex}
\centering
\includegraphics[width=0.45\textwidth]{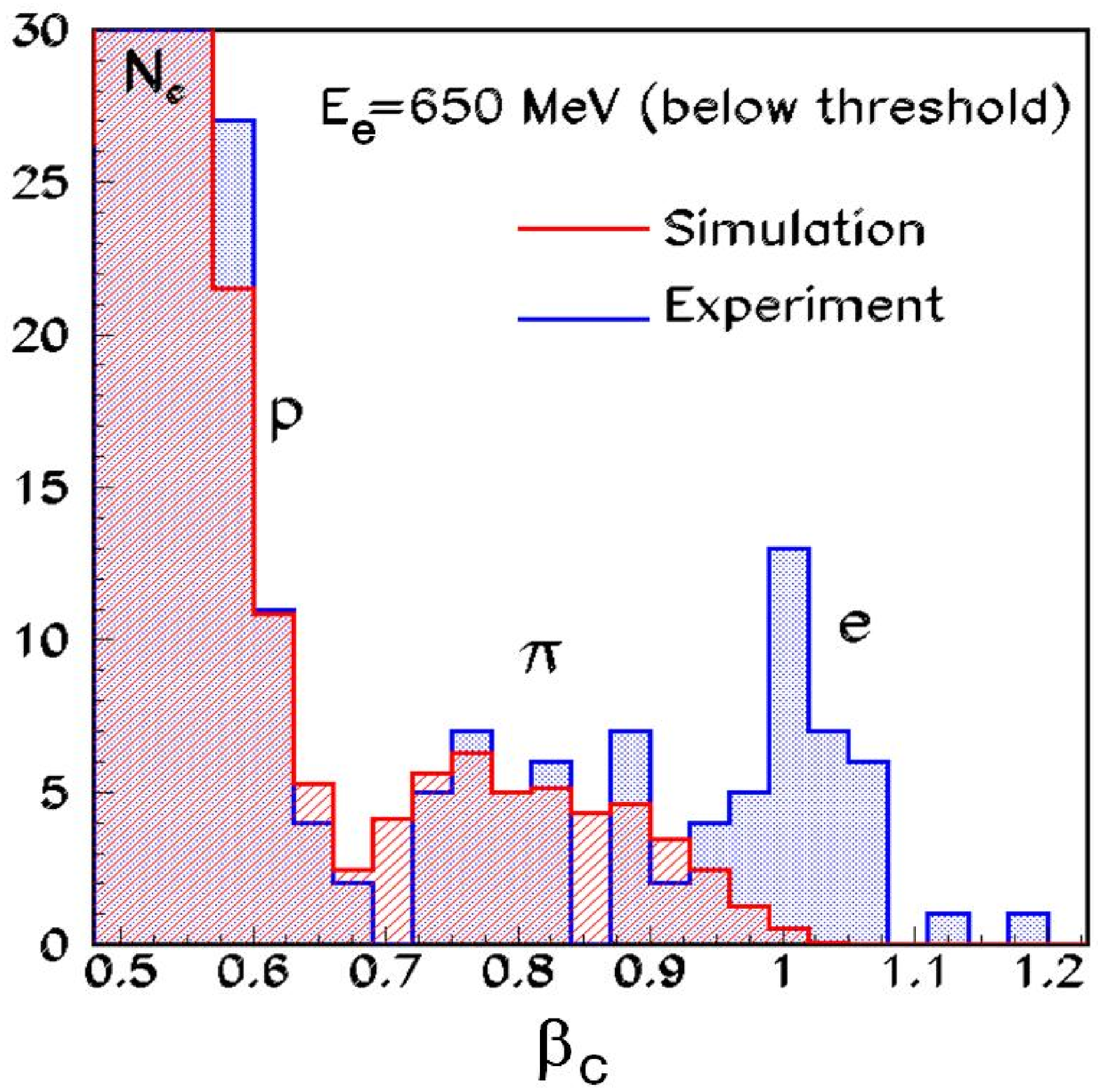}
\hspace{5ex}
\raisebox{0.5ex}{\includegraphics[width=0.445\textwidth]{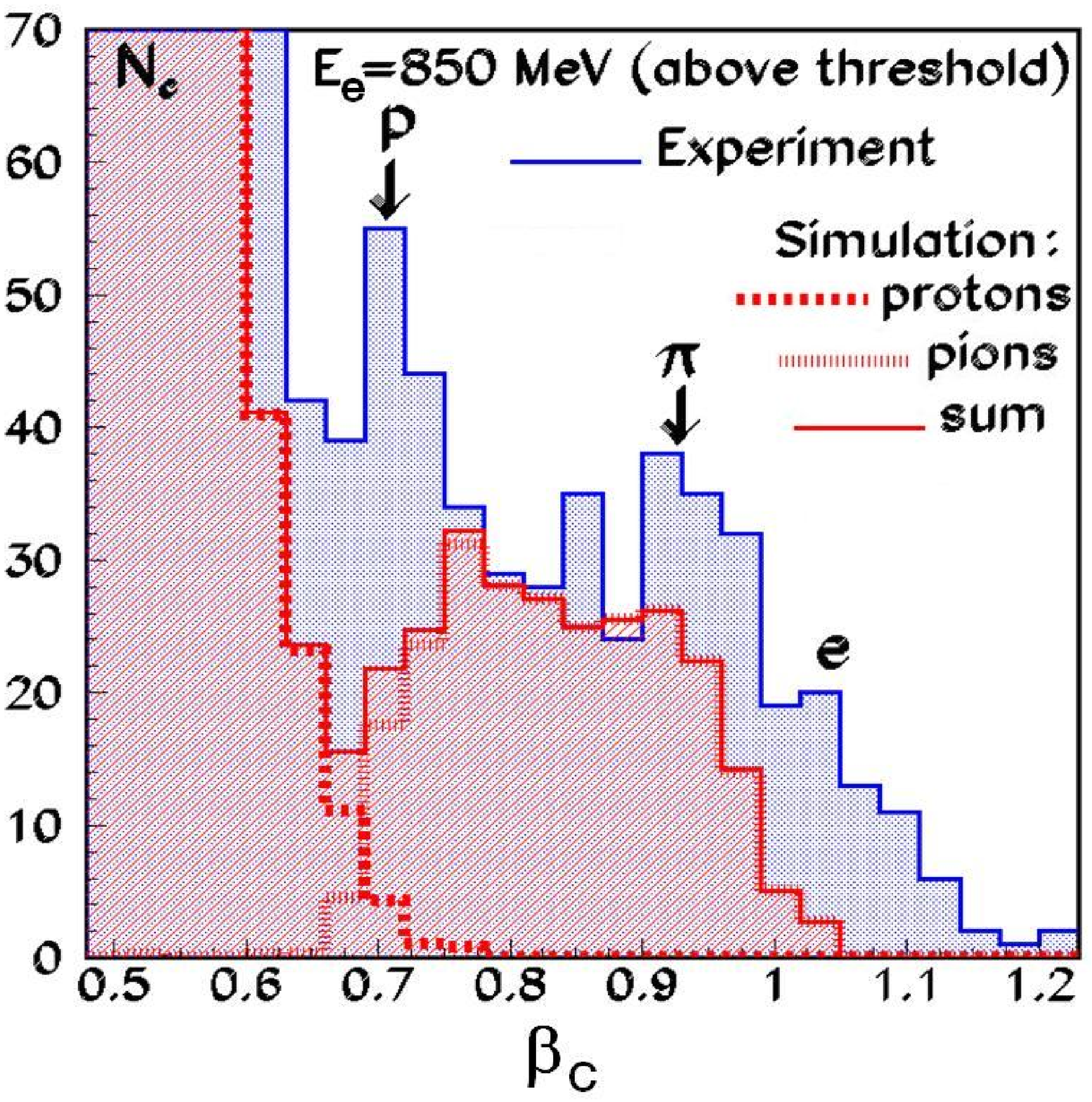}}
\vspace{-2ex}
\caption{Velocity distribution of charged particles selected according
to the criterion $\Delta E_i > 0$ (for all $i=1,2,3$) at $E_e=650$ and
850 MeV. A well visible excess of events over the INC simulation is seen
at the right panel --- in the case of the beam energy exceeding the
$\eta$-photoproduction threshold --- in both velocity regions
corresponding to the expected velocities of the $\pi N$ and $NN$ decay
products of $\eta$-mesic nuclei.}
\label{bC}
\vspace{3ex}
\end{figure}

Considering one-dimensional spectra over $\beta_C$ of events selected
according to the condition (\ref{eq:selection}) of sufficiently long
ranges and imposing the additional cut-off $0.3 <\beta_N < 0.7$ for
neutron velocities, we find rather interesting structures in the
spectra. Shown in Fig.~\ref{bC} are experimental data (blue areas)
together with results of the INC simulation (pink hatched areas).
Separately shown are INC predictions for the number of protons and
charged pions in the C-arm. There is a qualitative agreement of the
INC simulation with the experimental data for the case of the
subthreshold beam energy, $E_e=650$ MeV. Meanwhile, in the case of
$E_e=850$ MeV there is a clear excess of the experimentally observed
events over the simulation results in two velocity regions closely
corresponding to the kinematics of $\eta$-mesic nuclei decays with
emission of $\pi N$ and $NN$ correlated pairs, Eqs.~(\ref{kinema-piN})
and (\ref{kinema-NN}).

Knowing from the INC simulations that the ''normal'' (without
$\eta$-mesic nuclei) dynamics of the considered reaction does not
yield a sufficient amount of protons and pions with the velocities of
about $\beta_C\sim 0.7$, we interpret the found anomaly at
$\beta_C\sim 0.7$ as a result of production of low-energy
$\eta$-mesons followed by their two-nucleon annihilation.

The energy resolution of the experimental setup is not sufficient to
say whether an essential part of these $\eta$-mesons is produced in
the bound state, but theoretical arguments discussed in above make
such a statement plausible.

Concerning the excess of pions with $\beta_C\simeq 0.95$, this feature
is in agreement with our measurements reported earlier
\cite{sokol99,sokol00,sokol08}. It can be interpreted as an evidence
for one-nucleon annihilation of produced low-energy $\eta$-mesons
(bound or unbound).

Electron/positron peaks shown in Fig.~\ref{bC} originate from
calibration runs with the lead plate inserted. They were not included
into simulations made.

The observed proton peak in the $\beta_C$ distribution is very unusual
because it corresponds to $pn$ pairs with very high kinetic energies
$T_p\sim T_n\sim 200{-}300$ MeV and transverse momenta $p_p\sim
p_n\sim 400{-}800$ MeV/c. One should keep in mind that photons which
produce such pairs have quite a modest energy $650~{\rm MeV} < E_\gamma
< 850$ MeV. Ordinary photoproduction reactions do not give nucleons
with such a high energy and momentum. Creation and annihilation of
intermediate low-energy $\eta$-mesons seems to be the only explanation
to these events.

Assuming that the observed access events are mainly related with
formation and isotropic decays of $\eta$-mesic nuclei with $A=11$, we
can estimate their photoproduction cross section. The number of
photons of the energies $E_\gamma = 650{-}850$ MeV that hit the carbon
target in experimental runs was evaluated via comparison of the total
yield of charged pions detected by a single C-arm of the setup with
predictions of INC for that yield, thus giving the result $N_\gamma
\simeq 1.36\times 10^{11}$. Taking into account the solid angle of the
C-arm telescope ($\Omega_C = 0.027$~sr), efficiencies of detectors, a
geometric efficiency of the $N$-arm of the setup ($\sim 18\%$ as found
from theoretically expected angular distributions of particles of the
correlated pairs), we arrived at the following cross section of
$\eta$-mesic nucleus formation:
\beq
  \sigma(\gamma + {~}^{12}{\rm C}\to{}_\eta A + X) \lesssim 10~\rm \mu b.
\label{x-section}
\eeq
We write it as an upper limit because part of the observed events can
be related with unbound etas. This number is consistent with available
theoretical estimates (typically, a few $\mu$b).

\section*{Conclusions}

The new obtained data confirm the main features of the $\pi N$ signal
of formation and decay of $\eta$-mesic nuclei off the carbon target in
the photoreaction found in our previous work.

A new signature for formation and decay of $\eta$-mesic nuclei, the
back-to-back $pn$ pairs, was explored. For the first time an
experimental evidence was found that the yield of such pairs in the
region of $\beta_C\sim 0.6{-}0.7$ is quite large and therefore is also
suitable for searching for $\eta$--mesic nuclei.

Assuming that the observed excess of events is related with
$\eta$-mesic nuclei, an estimate of the total cross section of
formation of $\eta$-nuclei in the photoreaction off carbon have been
obtained, see Eq.~(\ref{x-section}).

We have plans to carry out a more precise experiment, with a better
energy resolution, at the deuteron beam of the JINR nuclotron.

\section*{Acknowledgments}

This work was supported in part by the RFBR grants 08-02-00648-a and 10-02-01433-a.
A nice work of the accelerator group of the LPI synchrotron and its
leader G.G.~Subbotin is highly appreciated.

\end{document}